\documentclass[english,aps,prX,showpacs,twocolumn]{revtex4-1}
\usepackage[T1]{fontenc}
\usepackage[latin9]{inputenc}
\usepackage{babel}
\usepackage{graphicx}
\usepackage{bm}
\usepackage{bbm}
\usepackage{array}
\usepackage{amsmath}
\usepackage{amssymb}
\usepackage{dcolumn}
\usepackage{epstopdf}
\usepackage{array}
\usepackage{tabu}
\usepackage{bbold}
\usepackage{tikz}
\usepackage[margin=15mm]{geometry}
\colorlet{linkequation}{blue}
\usepackage[colorlinks=true,linkcolor=blue,citecolor=blue,urlcolor=blue]{hyperref}
\usetikzlibrary{shadings}

\usepackage{color}
\newcommand{\blue}{\color{blue}}

\begin{document}

\title{Unified theory of magnetization dynamics with relativistic and nonrelativistic spin torques}

\author{Ritwik Mondal}
\altaffiliation{Present address: Fachbereich Physik, Universit\"at Konstanz, DE-78457 Konstanz, Germany}
\email[]{ritwik.mondal@uni-konstanz.de}

\author{Marco Berritta}
\affiliation{Department of Physics and Astronomy, Uppsala University, P.\,O.\ Box 516, Uppsala, SE-75120, Sweden}

\author{Peter M. Oppeneer}
\affiliation{Department of Physics and Astronomy, Uppsala University, P.\,O.\ Box 516, Uppsala, SE-75120, Sweden}

\date{\today}

\begin{abstract}

Spin torques play a crucial role in operative properties of modern spintronic devices. 
To study current-driven magnetization dynamics, spin-torque terms providing the action of spin-polarized currents have previously {often} been added in a phenomenological way to the Landau-Lifshitz-Gilbert equation describing the local spin dynamics, yet without derivation from fundamental principles.
Here, starting from the Dirac-Kohn-Sham theory and incorporating nonlocal spin transport we rigorously derive the various spin-torque terms that appear in current-driven magnetization dynamics. 
In particular we obtain an extended magnetization dynamics equation that precisely contains the nonrelativistic adiabatic and relativistic nonadiabatic spin-transfer torques (STTs) of the Berger and Zhang-Li forms as well as relativistic spin-orbit torques (SOTs). 
We derive in addition a previously unnoticed relativistic spin-torque term and moreover show that the various obtained spin-torque terms do not appear in the same mathematical form in both the Landau-Lifshitz and Landau-Lifshitz-Gilbert equations of spin dynamics.

\end{abstract}



\maketitle

\section{Introduction}
Along with the traditional Landau-Lifshitz-Gilbert (LLG) \cite{Landau1935,Gilbert1956,Gilbert2004} equation of motion of magnetization dynamics, there exist additional spin torques that have  emerged as a new way to manipulate spins in spintronics \cite{Wolf2001,Kent2015} and spin-orbitronics \cite{Kuschel2015}. These torques stem from the spin-polarized current density flowing in a magnetic material. There are two kinds of torques that have been identified in experiments: (1) spin-transfer torques (STTs) that arise due to the one-directional current flow perpendicular to the sample's (interface) plane. As a  consequence angular momentum is exchanged between two layers  that have different spin orientations. These torques have been instrumental to describe the domain wall motion of heterostructured magnetic layers
\cite{Slonczewski1996,Berger1996,Stiles2002,Zhang2004,Ralph2008} and ultrafast magnetization dynamics \cite{Kim2016,Cheng2015}. 
The spin-transfer torques are due to a nonrelativistic effect which is reflected in the fact that the current density is given by the nonrelativistic velocity operator. There exists also (2) the spin-orbit torque (SOT), {whose origin} is rather different {from that of} the STT,  and, in addition, it acts in the plane of the magnetic film \cite{Gambardella2011}. It is a relativistic effect where the angular momentum is exchanged between the crystal lattice and the spin degrees of freedom thus enabling to control the magnetic state even in a single magnetic layer. The spin-orbit interaction produces a current which strongly depends on the magnetization and the electric field. As the magnetization precesses around an effective magnetic field, the direction of the magnetization changes and thus the current generated from spin-orbit interaction is not unidirectional unlike the STT, but rather it remains in a plane depending on the magnetization. Recently, there have been successful attempts to manipulate magnetization through SOTs in  magnetic heterostructures \cite{Ando2008,Miron2011,Liu2012a,Liu2012b,Garello2013,Yu2014,Safeer2016,Baumgartner2017}, in ferromagnet-antiferromagnet bilayers \cite{Fan2014},  {antiferromagnets \cite{Wadley2016},} magnetic-doped topological insulators \cite{Fukami2016}, fast spin switching \cite{Garello2014}, and femtosecond control of magnetization in ferromagnets \cite{Huisman2016}. 

To simulate the effect of spin torques on the ensuing magnetization dynamics in a magnetic system, the spin torques have often simply been added to the LLG equation,
\begin{equation}
\frac{\partial \bm{M}}{\partial t} = -\gamma \,\bm{M} \times \bm{H}_{\rm eff} + \alpha \,\bm{M} \times \frac{\partial \bm M}{\partial t} + \bm{T},
\label{LLG-T}
\end{equation}
where $\bm{M}(\bm{r},t)$ is a local magnetization, $\bm{H}_{\rm eff}$ the effective magnetic field, $\gamma$ the gyromagnetic ratio, $\alpha$ the Gilbert damping parameter, and $\bm{T}$ the spin torque.
Despite the fact that the LLG equation is intensively used in spin-dynamics simulations (see, e.g., \cite{Brown1963,Nowak2007,Evans2014,Ellis2015}), it is a phenomenological equation proposed originally on the basis of physical intuition of the pioneers in the field \cite{Landau1935,Gilbert1956,Baryakhtar2015}. A derivation of Eq.\ (\ref{LLG-T}) from fundamental principles is thus still sollicitated.

On the theoretical side, explicit forms of the spin torque $\bm{T}$ have been investigated (see, e.g., \cite{Zhang2002,Zhang2004,Shpiro2003,Tatara2008PRB,Brataas2012,Kim2015}). It is known that spin torques arise from the nonequilibrium spin density \cite{Zhang2004}.
The common procedure to derive them is to consider an interaction Hamiltonian of conduction electrons and a local magnetization in the ``\textit{s-d}'' exchange Hamiltonian form $\bm{s}\cdot\bm{M}$, where $\bm{s}$ and $\bm{M}$ denote the itinerant electrons and local magnetization, respectively. The torque is then proportional to $\bm{m}\times \bm{M}$, with $\bm{m}(\bm{r},t) = \langle \bm{s} \rangle$  the itinerant spin density. The nonequilibrium spin density, denoted as $\delta \bm{m}$, exerts the torque $\delta \bm{m}\times\bm{M}$, which is the origin of the spin-transfer torque in ferromagnets \cite{Zhang2004,Kim2015,Tatara2008PRB,Shpiro2003}, ferromagnetic semiconductors \cite{Culcer2009}, and Weyl semimetals \cite{Nomura2015}. Although these derivations have been well understood, they provide an explicit form of the spin-transfer torque only. The spin-orbit torque is fundamentally more difficult, because it depends on a small relativistic term. To derive explicit expressions for the SOT often a particular form, such as Rashba spin-orbit interaction \cite{Manchon2015}, has been considered (see, e.g., \cite{Manchon2008,Manchon2009,Kim2013}). A derivation of the spin equation of motion with spin torques on the basis of relativistic electronic structure theory has not yet been given.

Recently, we have shown that the LLG equation can be fully derived in a relativistic Dirac-Kohn-Sham framework, leading to an explicit expression for the Gilbert damping \cite{Mondal2016}. Apart from the Gilbert damping, previously other types of damping, such as the spin nutation or magnetic inertia have been included in an extended LLG equation of motion through a second order time-derivative of the magnetization \cite{Ciornei2011,Fahnle2011,Bottcher2012}.
We have shown lately that this inertial damping torque can also be fully derived in the Dirac-Kohn-Sham framework and arises from higher-order spin-orbit coupling terms which act on ultrashort timescales only \cite{Mondal2017Nutation}. Along the same direction, we have provided a  generalized theory for spin dynamics, wherein we derived the most general form of the intrinsic Gilbert damping and magnetic inertia parameters from relativistic Dirac theory \cite{MondalJPCM2018}. The derived forms suggest that both these parameters can be expressed as a series of higher-order relativistic corrections and that they are tensors. 
 It deserves to be mentioned that there exist still other variants of the LLG equations. Bar'yakhtar {\it et al.}\ introduced an exchange damping torque, as a second order spatial derivative of the exchange field \cite{Baryakhtar1984,Baryakhtar2013}. To describe longitudinal spin relaxation the Landau-Lifshitz-Bloch equation has been derived \cite{Garanin1991,Chubykalo2006}. The effect of transverse spin diffusion has  furthermore been considered in the case of ferromagnets within the LLG framework \cite{Tserkovnyak2009}. Another recent work also proposed the existence of a torque which involves both spatial and temporal derivatives of the magnetization \cite{Li2016PRL}.

In this article we aim to derive the current-driven magnetization dynamics from fundamental principles. To this end we start from the Dirac-Kohn-Sham {(DKS)} Hamiltonian and first derive, in the semirelativistic limit, all spin-related nonrelativistic and relativistic Hamiltonian terms using an unitary transformation. Next, we proceed to the spin dynamics including spin-polarized currents, utilizing a continuity equation, which involves the spin density and then rigorously
calculate the spin torques. 
We observe two fundamentally different dynamics, the nonrelativistic magnetization dynamics, derived solely from the nonrelativistic Hamiltonian, and the relativistic magnetization dynamics which mainly is related to the spin-orbit coupling. Our derivation leads to the current-induced spin-transfer torque of Berger type \cite{Berger1996}, which is a nonrelativistic effect. The Zhang-Li nonadiabatic spin-transfer torque \cite{Zhang2004} is obtained when the spin-dynamics equation is rewritten in the Landau-Lifshitz (LL) form. The derivation of the relativistic spin dynamics contrarily leads to various electric field-induced torques, which are essentially spin-orbit torques.   

In the following we first introduce the relativistic DKS Hamiltonian with external fields. The magnetization dynamics with currents and external electric field is derived in Sec.\ III and discussed in Sec.\ IV, and our conclusions are given in Sec.\ V.

\section{Relativistic Hamiltonian Formulation}
We start from the Dirac-Kohn-Sham Hamiltonian which describes a relativistic spin-$\frac{1}{2}$ quantum particle efficiently in the effective mean-field formulation. For a magnetic material the DKS Hamiltonian is given as \cite{Macdonald1979,Eschrig1999,Greiner2000,Mondal2016,Kraft1995}
\begin{eqnarray}
\mathcal{H}_{\rm DKS}&=&c\,\bm{\underline{\alpha}}\cdot\left(\bm{p}-e\bm{A}\right)+
\left(\underline{\beta}-\mathbb{\underline{1}}\right)mc^{2}+ 
V \mathbb{\underline{1}} + e\Phi \mathbb{\underline{1}} \nonumber \\
& &- \mu_B \underline{\beta} \, \bm{\underline{\Sigma}}\cdot\bm{B}^{\rm xc}\,,
\label{eq:DKSH_xc}
\end{eqnarray} 
where $V$ stands for the unpolarized Kohn-Sham selfconsistent potential, 
$\bm{B}^{\rm xc}$ defines the spin-polarized exchange-correlation potential in the material, $\bm{A} = \bm{A}(\bm{r},t)$ is the magnetic vector potential due to an applied electromagnetic field (e.g., light), $e\Phi (\bm{r},t)$ is the scalar potential of this field, $\bm{p}=-i\hbar\bm{\nabla}$, the momentum operator  and 
$\mu_B =$$ \frac{e \hbar}{2m}$, the Bohr magneton.
$\underline{\mathbb{1}}$ is the $4\times4$ identity matrix and  $\underline{\bm{\alpha}}$, ${\underline{\beta}}$, and $\underline{\bm{\Sigma}}$
are the well-known Dirac matrices defined e.g., in Ref.\ \cite{Strange1998}.
Evidently there are two fundamentally different fields present in the DKS Hamiltonian, one is the Maxwell fields from the externally applied source and the other one is exchange field which couples to the spin degrees of freedom and has a very different origin. Consequently, the exchange field does not fulfill the Maxwell equations and can therefore not be treated as a minimal coupling to the momentum, i.e., $\bm{p}-e\bm{A}^{\rm xc}$ \cite{Eschrig1999,mondal15}.

Next, we want to investigate the relativistic spin dynamics of spin-polarized electrons including the effect of spin currents. To achieve this we need the full Pauli Hamiltonian for electrons in a magnetic system, including all the relativistic corrections that stem from the DKS Hamiltonian. To reach this we employ the Foldy-Wouthuysen (FW) transformation \cite{Foldy1950,Strange1998} on the DKS Hamiltonian. The FW transformation is a canonical and unitary transformation which is iteratively applied \cite{Silenko2016PRA}. The main idea of the iterative transformation is to make the off-diagonal (i.e., particle--antiparticle) elements smaller until the separation of particle and antiparticle is guaranteed for any given momentum \cite{Foldy1950}.  
This leads to a semirelativistic, extended Pauli (EP) Hamiltonian, $\mathcal{H}_{\rm EP}$, which includes all spin-dependent interactions and relativistic corrections as well as the exchange field and external electromagnetic fields \cite{mondal15,Mondal2016}. For convenience sake, we explicitly separate the extended Pauli-like Hamiltonian in nonrelativistic and relativistic Hamiltonian terms,  i.e., $\mathcal{H}_{\rm EP} = \mathcal{H}_{\rm NR} + \mathcal{H}_{\rm R}$, which are defined as \cite{Mondalthesis}
\begin{widetext}
\begin{align}
\label{FW_hamiltonian-NR}
\mathcal{H}_{\rm NR} =~ & \frac{\left(\bm{p}-e\bm{A}\right)^{2}}{2m}+V  -  \mu_B \,\bm{\sigma}\cdot \left(\bm{B} + \bm{B}^{\rm xc}\right)  +
e\Phi \\
\label{FW_hamiltonian-R}
\mathcal{H}_{\rm R}  =~ &- \mu_B \bm{\sigma} \cdot \bm{B}^{\rm xc}_{\rm corr} - \frac{\left(\bm{p}-e\bm{A}\right)^{4}}{8m^{3}c^{2}}-
\frac{e\hbar^{2}}{8m^{2}c^{2}}\bm{\nabla}\cdot\bm{E}_{\rm tot} -\frac{e\hbar}{8m^{2}c^{2}}\bm{\sigma}\cdot\left[ \bm{E}_{\rm tot}\times\left(\bm{p}-e\bm{A}\right)-\left(\bm{p}-e\bm{A}\right)\times\bm{E}_{\rm tot}\right]\nonumber\\
& +\frac{i \mu_B}{4 m^{2}c^{2}}[(\bm{p}\times\bm{B}^{\rm xc})\cdot\left(\bm{p}-e\bm{A}\right)] \,.
\end{align}
 All the appearing relativistic corrections involving the exchange interaction  (with exception of the last term in Eq.\ (\ref{FW_hamiltonian-R})) have here been added together to give the relativistic \textit{correction} to the exchange field \cite{note1}, 
 \begin{eqnarray}
\bm{B}^{\rm xc}_{\rm corr} & = &   - \frac{1}{8m^{2}c^{2}}\Big\{ \! \left[p^{2}\bm{B}^{\rm xc}\right]+2(\bm{p}\bm{B}^{\rm xc})\! \cdot \! \left(\bm{p}-e\bm{A}\right)+2(\bm{p}\cdot\bm{B}^{\rm xc})\!\left(\bm{p}-e\bm{A}\right)+4[\bm{B}^{\rm xc}\! \cdot \! \left(\bm{p}-e\bm{A}\right)] \!\left(\bm{p}-e\bm{A}\right)\Big\}\,.
\label{Bxc-eff}
\end{eqnarray}
\end{widetext}
Thus, the exchange fields appearing in the DKS Hamiltonian ($\bm{B}^{\rm xc}$) and in the extended Pauli-like Hamiltonian ($\bm{B}^{\rm xc} +  \bm{B}^{\rm xc}_{\rm corr}$) are not identical.
The other fields that are present in the above-derived relativistic Hamiltonian $\mathcal{H}_{\rm EP}$ are: the external magnetic field, $\bm{B} (\bm{r}, t) =\bm{\nabla}\times\bm{A} (\bm{r},t) $ as well as the total electric field that is the sum of internal, $\bm{E}_{\rm int}$ and external, $\bm{E}_{\rm ext}$ contributions as $\bm{E}_{\rm tot} = \bm{E}_{\rm int} + \bm{E}_{\rm ext}$. 
The internal electric field stems from the ions and the interactions of electrons, which even exists without any perturbation i.e., $\bm{E}_{\rm int} = -\frac{1}{e}\bm{\nabla}V$. On the other hand the external field is defined as $\bm{E}_{\rm ext} = -\frac{\partial \bm{A} }{\partial t} - \bm{\nabla}\Phi$. 
 The meaning of each of the terms in the Hamiltonians (\ref{FW_hamiltonian-NR}) and (\ref{FW_hamiltonian-R})  are immediately explained, see Ref.\ \cite{mondal15} for details.

\section{Magnetization dynamics}
Now that we have all the required terms in the EP Hamiltonian we can proceed to derive the full magnetization dynamics with spin-current terms included.

The magnetization is, according to its definition, given by the expectation value of the spin density \cite{White2007} 
\begin{align}
\label{magn}
{\bm{M}}(\bm{r},t) & = g\mu_B \big\langle \hat{\bm{S}}\, \delta(\bm{r}- \hat{\bm{r}})\big\rangle,
\end{align}
where $\mu_B=e\hbar/2m$ is the usual Bohr magneton and the Land\'e $g$ factor, $g\approx 2$ for spin degrees of freedom. 
Note that we have defined magnetization as spin angular momentum per unit volume in our earlier investigations \cite{Mondal2016,Mondal2017Nutation}, which is valid in the case of homogeneous spin distributions. However, for \textit{inhomogeneous} spin distributions the individual spins have to be considered,
 as is done in Eq.\ (\ref{magn}). Note that we do not consider the magnetization contribution due to the orbital moment,
because this quantity is quenched for the common transition metals  (e.g., Fe, Ni, Co etc.).
The equation of motion of the magnetization is obtained by taking the time derivative on both sides of Eq.\ (\ref{magn}), which gives
\begin{align}
\frac{\partial \bm{M}}{\partial t} & = g\mu_B \Big\langle\frac{\partial \hat{\bm{S}}}{\partial t}\delta(\bm{r}- \hat{\bm{r}})+\hat{\bm{S}}\frac{\partial\delta(\bm{r}- \hat{\bm{r}})}{\partial t}\Big\rangle\, .
\label{eq:motion1}
\end{align}
Next we use the Heisenberg representation of operators and the corresponding dynamical equation of motion. This leads to the full magnetization dynamics 
\begin{align}
\label{eq:motion2}
	\frac{\partial \bm{M}}{\partial t} & = \frac{g\mu_B}{i\hbar}  \Big\langle\big[\hat{\bm{S}},\mathcal{H}\big]\delta(\bm{r}- \hat{\bm{r}})+\hat{\bm{S}}\left[\delta(\bm{r}- \hat{\bm{r}}),\mathcal{H}\right]\Big\rangle\,.
\end{align} 
For the first commutator of spin angular momentum with the Hamiltonian $[\hat{\bm{S}},\mathcal{H} ]$ to be nonzero, the Hamiltonian must depend on the spin as otherwise the commutator vanishes. This commutator has been considered and worked out in an earlier study \cite{Mondal2016}, where we have already discussed the origin of magnetization precession and of Gilbert damping that follow completely from the first term in Eq.\ (\ref{eq:motion2}). All these phenomena are directly related to the local magnetization dynamics.
The second commutator, in contrast, involves the density operator and the Hamiltonian, which in turn defines the orbital current density. The latter will contribute to the \textit{nonlocal} processes in the magnetization dynamics, as e.g., current-induced magnetization dynamics, spin-transfer torque, spin-orbit torques etc.      

In the following we derive rigorously the effect of the second term on the right-hand side of Eq.\ (\ref{eq:motion2}) within the magnetization dynamics. 
For an electron, the charge density operator will be given as $\hat{\rho} = e\,\delta(\bm{r}- \hat{\bm{r}})$ \cite{Cho2008}. From the continuity equation we get $\partial\hat{\rho}/ \partial t = - \bm{\nabla}\cdot\hat{\bm{j}}$, where $\hat{\bm{j}}$ is the current density operator which has to be obtained from the Hamiltonians Eqs.\ (\ref{FW_hamiltonian-NR}) and (\ref{FW_hamiltonian-R}). Using these considerations together with the fact that the spatial derivative of spin angular momentum is zero, we arrive at \cite{Tatara2008PRB,Capelle2001PRL,Shi2006PRL}      
\begin{align}
\label{eq:motion_full}
\frac{\partial \bm{M}}{\partial t} + \bm{\nabla}\cdot J_{\rm S} = \bm{T}\,,
\end{align}
where $J_{\rm S}$ is the spin current and $\bm{T}$ is the torque acting on the spins that originates from the commutator of spin angular momentum with the Hamiltonian in Eq.\ (\ref{eq:motion2}). Note that this continuity equation is different from the usual ones because of the appearing torque term. We also note that spin relaxation processes due to the electron-phonon scattering have not been included in the present study. 
Incorporating scattering processes such as spin relaxation to the lattice or spin-defect scattering would give rise to nonadiabatic effects \cite{Zhang2004}. 
The {DKS} spin-current tensor is defined as
\begin{align}
J_{\rm S} &=  \frac{g\mu_B}{e} \Big\langle\hat{\bm{j}} \otimes \hat{\bm{S}} \Big\rangle = \frac{1}{e}\,\bm{j}\otimes\bm{M}\,.
\end{align}
This tensor depends on the charge current density $\bm{j}$ and the magnetization. 

To obtain expressions for the spin-current terms we need to calculate the terms in Eq.\ (\ref{eq:motion2}).
At this point it is important to note that neither the spin nor the orbital degrees of freedom commute with the Hamiltonian due to the spin-orbit coupling and  the  other $1/c^2$ corrections.  This in turn implies that the equilibrium density operator $\hat{\rho}$
cannot be expressed exactly as $\hat{\rho}=\hat{\rho}_S\otimes\hat{\rho}_O$ where $\hat{\rho}_S$ is the reduced density  operator for the spin degrees of freedom and $\hat{\rho}_O$ the reduced density operator of the orbital degrees of freedom. Considering  an observable ${O}$ acting on the orbital degrees of freedom in Hilbert space (for instance the momentum or the orbital angular momentum or some function depending on them) and $\hat{{\bm S}}$ the spin, then, due to the impossibility to separate orbital and spin parts of the density matrix, we are in principle not allowed to write 
\begin{equation}
\label{eq:sep}
{\rm Tr}[\hat{\rho}\hat{{\bm S}}{O}]={\rm Tr}[\hat{\rho}_S \hat{{\bm S}}]\,{\rm Tr}[\hat{\rho}_O {O}]={\bm M}\langle{O}\rangle. 
\end{equation}
To nonetheless be able to employ this approximation it is important to realize that the nonseparability (entanglement) of the orbital and spin parts of the density matrix is due to the spin-orbit coupling and its corrections (since it prevents each of these quantities to be conserved). However, in ferromagnetic materials the energy separation of the spin states is mostly due to the exchange field which is orders of magnitude larger than the spin-orbit coupling and its corrections. As a consequence, the separation of the density operator as a direct product of spin and orbital parts is a good approximation, and therefore we can employ Eq.\ (\ref{eq:sep}). Moreover, due to the large splitting of spin bands and the continuous smooth behavior of the energy levels as function of momentum, the out-of-equilibrium dynamics on the latter degrees of freedom is much faster than the dynamics of the spin degrees of freedom.

The tensor form of the spin current has been discussed earlier in the context of time-dependent spin-density functional theory \cite{Sun2005PRB,Capelle2001PRL,An2012}. 
In the nonrelativistic Kohn-Sham Hamiltonian the current density stems from the commutator of kinetic energy with the particle density. However, the picture is different when we consider the relativistic {DKS} formalism. 

The associated relativistic current density can be derived from the usual prescription of Landau and Lifshitz \cite{Landau1981} in which it is given by the variation of interaction energy with respect to the magnetic vector potential as \cite{mondal15,Hinschberger2013}
\begin{eqnarray}
\delta\langle\mathcal{H}_{\rm int}\rangle=- e \int d\bm{r}\, \bm{j} \cdot \delta\bm{A}\,.
\end{eqnarray}
As the interaction Hamiltonians in Eqs.\ (\ref{FW_hamiltonian-NR}) and (\ref{FW_hamiltonian-R}) contain nonrelativistic and relativistic parts, we can consequently decompose the current density contributions into a nonrelativistic current density $\bm{j}_{\rm NR}$ and a relativistic current density $\bm{j}_{\rm R}$. The total current density can then be expressed as $\bm{j} = \bm{j}_{\rm NR} + \bm{j}_{\rm R}$. Similarly, the corresponding torques have nonrelativistic and relativistic origins, and we can formally write the magnetization dynamics as:
\begin{align}
\label{nonrelativistic-dynamics}
\frac{\partial \bm{M}}{\partial t}\Big \vert_{\rm NR} + \frac{1}{e}\Big[\bm{M}\left( \bm{\nabla}\cdot \bm{j}_{\rm NR}\right) + \left(\bm{j}_{\rm NR} \cdot \bm{\nabla}\right)\bm{M}\Big] & = \bm{T}_{\rm NR}\,,\\
\label{relativistic-dynamics}
\frac{\partial \bm{M}}{\partial t}\Big \vert_{\rm R} + \frac{1}{e}\Big[\bm{M}\left( \bm{\nabla}\cdot \bm{j}_{\rm R}\right) + \left(\bm{j}_{\rm R} \cdot \bm{\nabla}\right)\bm{M} \Big] & = \bm{T}_{\rm R}\,.
\end{align}
{Note that this formulation is done to shorten the notation; the complete $\partial \bm{M} / \partial t $ dynamics is given below.
In Eqs.\ (\ref{nonrelativistic-dynamics}) and (\ref{relativistic-dynamics})} we have used that the divergence of a dyadic tensor product $\bm{A}\otimes\bm{B}$ is simply given by the following identity, $\bm{\nabla}\cdot(\bm{A}\otimes\bm{B})=(\bm{\nabla}\cdot\bm{A})\bm{B}
+(\bm{A}\cdot\bm{\nabla})\bm{B}$ for any two vectors $\bm{A}$ and $\bm{B}$. It is interesting to note that we can thus distinguish the dynamical equations for magnetization due to nonrelativistic and relativistic origins, however the same is not possible while defining magnetization only. 

In the following we calculate the relativistic and nonrelativistic dynamics.

\subsection{Nonrelativistic magnetization dynamics}

 The current density and torques that appear in the nonrelativistic part [Eq.\ (\ref{nonrelativistic-dynamics})] can be calculated from the nonrelativistic Hamiltonian Eq.\ (\ref{FW_hamiltonian-NR}), where the nonrelativistic part of the effective exchange field is $\bm{B}^{\rm xc}$ has to be considered.
Let $\psi$ be the spinor solution to the EP Hamiltonian. The associated current density then becomes \cite{Nowakowski1999,Hinschberger2013}  
\begin{align}
\label{NR-currnet}
\bm{j}_{\rm NR}(\bm{r})  = & \frac{i\hbar }{2m}\big(\psi \bm{\nabla}\psi^{\dagger}-\psi^{\dagger}\bm{\nabla}\psi\big)-\frac{e}{m}\bm{A}\psi^\dagger\psi\nonumber\\
& +\frac{1}{m}\bm{\nabla}\times\big(\psi^\dagger\bm{S}\psi\big)\,.
\end{align}
The first term derives from the commutator of the particle density with the kinetic energy $p^2/2m$, the second term is the current due to the external perturbation from the electromagnetic field, i.e., the term involving $\bm{A}\cdot\bm{p}$. The last term is derived from the ``Zeeman-like'' term in Eq.\ (\ref{nonrelativistic-dynamics}) and is due to the fact that the external magnetic field can be expressed within the Coulomb gauge as $\bm{B}=\bm{\nabla}\times\bm{A}$ \cite{Hinschberger2013,mondal15}. A detailed derivation of this magnetic current term can be found in Ref.\ \cite{mondal15}. 
As we mentioned before, the exchange field solely stems from the Pauli exclusion principle and depends on spin degrees of freedom only, 
hence, it cannot be treated as Maxwellian field \cite{Eschrig1999}. Thus, there is no contribution of current density coming from an exchange field.
Using the definition of magnetization, 
 the last term could be written in the form $\frac{1}{mg\mu_B}\bm{\nabla}\times\bm{M}$, which also defines the bound current density according to classical electrodynamics \cite{Jackson}. 

Along the same reasoning the nonrelativistic torques on the spins will stem from the commutator of spin angular momentum with the Hamiltonian. Understandably, only the ``Zeeman-like'' couplings with external and exchange fields will contribute to these torques. In the magnetization dynamics, these torques can be summed up and describe the precession of the magnetization vector around an effective  nonrelativistic field \cite{Mondal2016},
\begin{align}
	\bm{T}_{\rm NR} & = -\gamma_{0}\, \bm{M}\times\bm{H}_{\rm eff}^{\rm NR}\,,
\end{align}
where $\gamma_0 = \mu_0\gamma $ is the effective gyromagnetic ratio \cite{Lakshmanan2011} and the gyromagnetic ratio is defined as $\gamma = g\vert e \vert/2m$ \cite{Mondal2016}. Here,  the effective nonrelativistic field, $\bm{H}_{\rm eff}^{\rm NR}$, is due to the Zeeman coupling of external and exchange fields to the spins.
 Adiabatic spin-dynamics simulations have previously been developed for elemental ferromagnets on the basis of the nonrelativistic Kohn-Sham Hamiltonian \cite{Halilov1998}.

Now that we have all the required expressions we can proceed to analyze the magnetization dynamics given by Eq.\ (\ref{nonrelativistic-dynamics}).
Let us focus first on the the second term i.e., $\bm{M}(\bm{\nabla}\cdot\bm{j}_{\rm NR})$.
Due to the Coulomb gauge, $\bm{\nabla}\cdot\bm{A}=0$ and also the divergence of a curl is always zero, $\bm{\nabla}\cdot(\bm{\nabla}\times\bm{M})=0$. Thus, the external perturbation and the spin-polarized Zeeman current do not contribute to the dynamics and we are left with the first term in Eq.\ (\ref{NR-currnet}). Over a volume that contains the sample, the divergence of current density is zero as there is no source nor sink.
On the other hand, the third term in Eq.\ (\ref{nonrelativistic-dynamics}), i.e.\ $(\bm{j}_{\rm NR}\cdot\bm{\nabla})$, bears much importance for the magnetization dynamics as it treats the magnetization inhomogeneity and its spatial variation. 
Writing down the nonrelativistic dynamics, we have
\begin{align}
	\frac{\partial \bm{M}}{\partial t}\Big \vert_{\rm NR}  & = - \gamma_{0}\, \bm{M}\times\bm{H}_{\rm eff}^{\rm NR} - \frac{1}{e} \big(\bm{j}_{\rm NR} \cdot \bm{\nabla}\big)\bm{M}\,.
\end{align}
Interestingly, we have thus derived the Berger-type current-induced spin-transfer torque \cite{Berger1996}, which is expressed as $\bm{T}_{\rm STT} = \frac{1}{e}\big(\bm{j}_{\rm NR}\cdot\bm{\nabla}\big)\bm{M}$ in the adiabatic limit (see, e.g., Refs.\ \cite{Ansermet2004,Li2004,Thiaville2005, Ralph2008,Brataas2012,Cheng2013,Nomura2015,Kim2015}).
It is independent of the electron relaxation lifetime. Note that we have deliberately not taken into consideration the nonadiabatic current density; the latter would lead to the corresponding spin-transfer torque that essentially includes the spin-flip and scatterings processes \cite{Zhang2004}.

\subsection{Relativistic magnetization dynamics}

\begin{figure}[t!]
\centering
\includegraphics[width = 0.45\textwidth]{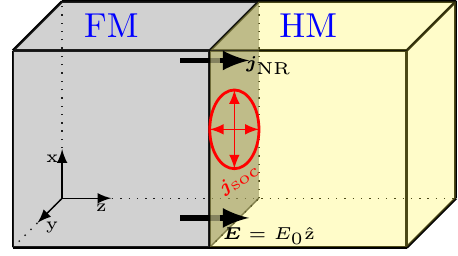}
\caption{Schematic illustration of the two different currents
{in a ferromagnet (FM)--heavy metal (HM) bilayer.} The nonrelativistic current $\bm{j}_{\rm NR}$ is unidirectional and provides a torque to the ferromagnetic spin  component that is perpendicular to the $z$-axis. In contrast, the relativistic spin-orbit current $\bm{j}_{\rm soc}$ lies in the $xy$-plane and provides a torque  on the spin which is along the $z$-direction.}
\label{soc-current}
\end{figure}

To obtain the relativistic current density the spin-orbit coupling and other relativistic effects need to be incorporated explicitly in the continuity equation. 
Thereto we write the current density as $\bm{j}_{\rm R}=\bm{j}_{\rm soc}+\bm{j}_{\rm other}$, where the latter term represents the other {(non-SOC)} relativistic current density. 
The main relativistic contribution to the current density comes from the spin-orbit coupling and it has the following form \cite{Hinschberger2013,mondal15} 
\begin{eqnarray}
\bm{j}_{\rm soc}(\bm{r})=-\frac{e}{2m^2c^2}\,\big(\psi^\dagger\bm{S}\psi\big)\times\bm{E}_{\rm tot} =  \xi \, \bm{M}\times\bm{E}_{\rm tot}\,,
\end{eqnarray}
where we defined $\xi=-\frac{e}{2m^2c^2g\mu_B}$. The direction and magnitude of this current depends on the magnetization and electric field. As the magnetization precesses, the direction of the current changes and thus generates a torque. 
 As mentioned earlier, the total electric field has two contributions, the internal and external field. {In turn,} the internal electric field has {also} two contributing components: (1) the internal electric field without any application of external field which we denote $\bm{E}^{0}_{\rm int}$, and (2) the response of the material with the applied external field, which in a linear approximation gives \cite{mondal15b}
\begin{eqnarray}
\bm{E}_{\rm int}=\bm{E}^{0}_{\rm int}+\Gamma \cdot \bm{E}_{\rm ext}\,.
\label{E-field}
\end{eqnarray}
In general $\Gamma$ is a material dependent tensor of rank 2 which reduces to a scalar for an isotropic material. For simplicity, in the following, and throughout this work we use $\Gamma$ as a scalar parameter. The zero-response internal electric field has the form $\bm{E}^{0}_{\rm int}=-\frac{1}{e}\bm{\nabla}V$. Thus, the current density due to the spin-orbit coupling will be given as
\begin{eqnarray}
\bm{j}_{\rm soc}(\bm{r})=  \xi\, \bm{M}\times\bm{E}_{\rm int}^0+\xi(1+\Gamma)\bm{M}\times\bm{E}_{\rm ext}\,.
\end{eqnarray}
Note that the first term here would exist even in the absence of any applied field. It accounts for the bound spin currents induced by the spin-orbit coupling.
The current density due to the spin-orbit coupling are at the heart of the anomalous Hall effect, where the off-diagonal conductivity is a function of the magnetization \cite{Husmann2006,Roman2009,Nagaosa2010ReviewAHE,Bellaiche2013AHE}. 
If we take a general $E$-field directed along the $z$-axis, the spin-orbit current stays in the plane perpendicular to the electric field, i.e., the $xy$ plane. This phenomenon is pictorially  shown in Fig.\ \ref{soc-current}.  Thus, this current plays an important role in the case of interfaces in a layered system of a ferromagnet and a heavy metal, where the spin-orbit strength is large \cite{Miron2011,Liu2012a,Liu2012b,Amin2016a,Amin2016b}. However, it is also interesting to note that this current can have a significant  impact on a plane for a single-layered system, when the spin-orbit coupling is large.   

Let us look at the details of the spin-orbit current in the case of ferromagnets. Consider a ferromagnet where the magnetization is in the $z$-direction i.e., $\bm{M} = M_z \hat{\bm{z}}$. The $x$ and $y$ components of the electric fields will contribute to the current because of the cross product. The spin-orbit current is then written as 
\begin{align}
	\bm{j}_{\rm soc} & = -\xi(1+\Gamma) M_z \left( E_y\hat{\bm{x}} - E_x\hat{\bm{y}}\right)\,.
\end{align}
Using that $\bm{j} = \bm{\sigma} \cdot \bm{E}$, the off-diagonal conductivity matrix elements can be written as $\sigma_{xy} = -\sigma_{yx} \propto -\xi (1+ \Gamma) M_z$, which represents the anomalous Hall effect. In this context, we mention that   
 $\xi(1+\Gamma)$ has been argued in Ref.\ \cite{Bellaiche2013AHE} to be the anomalous Hall conductivity.
 These relativistic current densities enter into the relativistic magnetization dynamics of Eq.\ (\ref{relativistic-dynamics}) as $\bm{M}\left( \bm{\nabla}\cdot \bm{j}_{\rm soc}\right)$  and  $\left(\bm{j}_{\rm soc} \cdot \bm{\nabla}\right)\bm{M}$.
 Unlike the nonrelativistic current density, as $\bm{j}_{\rm soc}$ depends on the magnetization texture, the divergence provides a dependence on the inhomogeneity in the magnetization. 
Calculating now the divergence of the current density due to spin-orbit coupling, we obtain
\begin{eqnarray}
\label{div-spin-orbit-current}
\bm{\nabla}\cdot\bm{j}_{\rm soc} &=& \xi\,\bm{\nabla}\cdot\big(\bm{M}\times\bm{E}_{\rm tot}\big)\nonumber\\
&=& \xi \Big[\bm{E}_{\rm tot}\cdot\big(\bm{\nabla}\times\bm{M}\big)-\bm{M}\cdot\big(\bm{\nabla}\times\bm{E}_{\rm tot}\big)\Big]\nonumber\\
&=& \xi \Big[\bm{E}_{\rm tot}\cdot\big(\bm{\nabla}\times\bm{M}\big) + (1+\Gamma)\bm{M}\cdot \frac{\partial \bm{B}}{\partial t}\Big]\,.
\end{eqnarray}
In the last step we have used {Faraday's law of induction,}  $\bm{\nabla}\times\bm{E}_{\rm ext} = -\partial\bm{B}/\partial t$, and that for the internal electric field the curl of a gradient is always zero.
The magnetic induction follows a linear relationship with magnetization such that $\bm{B} = \mu_0 \left(\bm{H} + \bm{M}\right)$ \cite{Mondal2016}.  Using this relation and making the  fundamental assumption of transversal spin motion, namely, that the magnitude of magnetization is conserved during the dynamics \cite{new-ref}, which leads to $\bm{M}\cdot\partial \bm{M}/\partial t = 0$, and we are left with only $\mu_0 \bm{M}\cdot \partial \bm{H}/\partial t$ for a general time-dependent field. However, for a harmonic external field, introducing a differential susceptibility $\chi$, the time-derivative leads to $\partial \bm{B}/ \partial t=\mu_0(1+\chi^{-1})\partial \bm{M}/ \partial t$ and the last term on the right-hand side of Eq.\ (\ref{div-spin-orbit-current}) vanishes. For simplicity we continue our analysis with the harmonic field in the following. {Results for a nonharmonic driving field are given in Appendix A.}

The other current density term in the relativistic magnetization dynamics will be simply given as $\xi\,\left[(\bm{M}\times\bm{E}_{\rm tot})\cdot\bm{\nabla}\right]\bm{M}$. On the same footing, the relativistic torques will be derived from the commutators of spin angular momentum with the Hamiltonian of spin-orbit coupling and other interaction terms related to the exchange fields. In a previous study we have already discussed that the other relativistic effects related to the exchange field create an effective field, $\bm{H}_{\rm eff}^{R}$ which is responsible for the relativistic precession \cite{Mondal2016}. We have also shown that the spin-orbit coupling related to the internal field, contributes to the relativistic precession as well \cite{Mondal2016}. However, the external spin-orbit coupling in the form of $\bm{\sigma}\cdot (\bm{E}_{\rm ext}\times \bm{p})$ contributes to the damping and relaxation processes, in particular, to the {intrinsic} Gilbert damping \cite{Hickey2009,Mondal2016}. The damping parameter $A$ that we have derived is a tensor, which includes a scalar Gilbert damping, an anisotropic Ising-like tensorial damping and an antisymmetric chiral  damping. For a harmonic field, the Gilbert damping tensor has two characteristically different contributions: one is electronic in nature and is given by an expectation value of the product of position and momentum operator as $\langle r_\alpha p_\beta \rangle$, and the other is magnetic, and given through the imaginary part of the susceptibility tensor \cite{Mondal2016,Mondal2017Nutation}. We remark here that for a general time-dependent magnetic field, the Gilbert damping involves an additional torque -- the field-derivative torque. {Its effect is to make the damping effectively frequency dependent. Also,} if the applied field very sharply increases, e.g., step-like, this term exerts a huge but instantaneous torque which might offer new ways to manipulate the spins. 
As we have obtained the field-derivative torque in the LLG form of the spin dynamics, we derive in Appendix A the form of this torque in the Landau-Lifshitz spin dynamics where it  leads to an additional term to the effective magnetic field. 

The other relativistic interaction Hamiltonian part, $\bm{\sigma}\cdot (\bm{E}_{\rm tot}\times \bm{A})$ is easily recognized as a gauge invariant part of the spin-orbit coupling. Without this term the spin-orbit coupling is not gauge invariant \cite{mondal15b}. For the internal electric field, this part of the Hamiltonian in Eq.\ (\ref{FW_hamiltonian-R}) vanishes, leaving thus only the contribution from the external field. The optical spin angular momentum density within paraxial approximation gives $\bm{\mathcal{J}}_{\rm s} = \epsilon_{0}\left( \bm{E}_{\rm ext}\times \bm{A}\right)$ \cite{Allen2003,Barnett2010}.  
The magnetization dynamics due to the interaction with the optical angular momentum can be written as
\begin{align} 
\frac{\partial \bm{M}}{\partial t}\Big\vert _{\rm opt} & =  - \gamma \bm{M}\times \bm{B}_{\rm opt}\,	,
\end{align}
where we have introduced the optomagnetic field that can be written as $\bm{B}_{\rm opt} = \frac{e}{2mc^2} \,(\bm{E}_{\rm ext}\times \bm{A})$ \cite{mondal15b,Mondal2016,Mondal2017JPCM}. The ensuing dynamics essentially defines the optical spin-orbit torque acting on the local magnetization. Note that the optomagnetic field is helicity dependent, thus possibly enabling helicity-dependent phenomena, as e.g.,  all-optical helicity-dependent magnetization switching \cite{Stanciu2007,Lambert2014,Berritta2016}. Several experimental evidences for the possible manipulation of spins by means of an optical spin-orbit torque have been reported  recently \cite{Nemec2012,Ramsay2015,Choi2017}. 

Taking together all these above-derived terms, the relativistic magnetization dynamics can be written as
\begin{align}
& \frac{\partial \bm{M}}{\partial t}\Big \vert_{\rm R} + \frac{\xi}{e}\Big[\bm{M}\left(\bm{E}_{\rm tot}\cdot (\bm{\nabla}\times \bm{M})\right) +  \left((\bm{M}\times\bm{E}_{\rm tot}) \cdot \bm{\nabla}\right)\bm{M} \Big] \nonumber\\
	& = - \gamma_0 \bm{M}\times \bm{H}_{\rm eff}^{R} + \bm{M}\times\left(A\cdot \frac{\partial\bm{M}}{\partial t}\right) -\gamma \bm{M}\times \bm{B}_{\rm opt}\,.
\end{align} 
This equation is one of the central results of this work because it derives the existence of field-induced SOTs  that are first order in the gradient of magnetization. We denote the field-induced spin-orbit torques as $\bm{T}_{\rm SOT}$, 
$ \bm{T}_{\rm SOT} = -\frac{\xi}{e} \left[\bm{M}\left(\bm{E}\cdot (\bm{\nabla}\times \bm{M})\right) + \left((\bm{M}\times\bm{E}) \cdot \bm{\nabla}\right)\bm{M}\right]$. We have dropped the subscript of the electric field and from now onwards, for convenience, we use $\bm{E}_{\rm tot} \equiv \bm{E}$.
It is important to remember that these electric field-induced torques can occur with the intrinsic as well as the externally applied field. 

Interestingly, the form of the second term in the $\bm{T}_{\rm SOT}$ has exactly the form that 
has been predicted using solely symmetry considerations \cite{Bijl2012}.  It is proportional to the electric field;  using $\bm{E}= \rho \cdot \bm{j}$ it can be rewritten in terms of 
the current density, where $\rho$ is the resistivity tensor.
The first term in the SOT, {$\bm{M} (\bm{E}\cdot (\bm{\nabla}\times \bm{M}) )$,} has to our knowledge not be reported before.
For the 2$^{\rm nd}$-term, let us briefly consider a simple case where the electric field is directed along the $z$-axis, $\bm{E} = \vert E_0\vert \hat{\bm{z}}$ (see Fig.\ \ref{soc-current}). This term then produces a torque which is proportional to  $E_z \left(M_y\frac{\partial}{\partial x} - M_x\frac{\partial}{\partial y}\right)\bm{M}$.  
 When the electric field is along a particular direction, this torque accounts for the inhomogeneity of the magnetization along the plane perpendicular to that $E$-field direction.
This torque resembles somewhat the Gilbert local damping term, but with the difference that this term is determined by the \textit{spatial} derivative of the magnetization. 

\subsection{Complete magnetization dynamics}
\label{sec:fulldyn}

The nonrelativistic magnetization dynamics produces the current-induced torque and the relativistic magnetization dynamics gives the $E$-field induced torques.  
Having all the terms we write the full magnetization dynamics as
\begin{align}
\label{eq:motion_full_final1}
\frac{\partial\bm{M}}{\partial t}  = & - \gamma_0\bm{M}\times\bm{H}_{\rm eff} +\bm{M}\times\Big( A \cdot\frac{\partial\bm{M}}{\partial t}\Big) \nonumber\\
& -\gamma \bm{M}\times \bm{B}_{\rm opt} - \frac{1}{e}\big(\bm{j}_{\rm NR}\cdot\bm{\nabla}\big)\bm{M} \nonumber\\
& -\frac{\xi}{e} \Big[\bm{M}\big(\bm{E}\cdot(\bm{\nabla}\times\bm{M}) \big)+\big((\bm{M}\times\bm{E})\cdot\bm{\nabla}\big)\bm{M}\Big]\,.
\end{align}
{This expression gives the field-driven magnetization dynamics of LLG form in a system with a nonuniform magnetization.}
The first two terms on the right-hand side have the usual physical interpretations. 
The third term defines the optical spin-orbit torque on the magnetization, related to a relativistic contribution to the inverse Faraday effect, where the optomagnetic field is proportional to $\bm{E}\times\bm{E}^\star$ i.e., intensity for a general electric field \cite{Mondal2017JPCM}.  
The fourth term is the Berger \cite{Berger1996} adiabatic STT as has already been pointed out before, and finally, the last {two} terms are the SOTs. 
As a side remark, we observe that Eq.\ (\ref{eq:motion_full_final1}) cannot be obtained from the usual LLG equation through substitution of convective derivates, i.e.\
$\frac{\partial}{\partial t} \rightarrow \frac{\partial }{\partial t} + \bm{v} \cdot \bm{\nabla} $, to account for the spin-polarized particle flow with velocity $\bm{v}$.

For practical purposes, it is often needed to write the spin dynamics in the form of the Landau-Lifshitz equation, where the $\partial \bm{M} / \partial t$ is eliminated from the Gilbert damping. For a scalar Gilbert damping parameter, the transformation is rather straightforward, and one obtains the LL equation that is mathematical equivalent to the LLG equation. This transformation is no longer straightforward for the case of tensorial Gilbert damping \cite{Mondal2016}. 
{The Gilbert damping tensor can be written as 
$ A_{ij}= \alpha \delta_{ij} + \mathbb{I}_{ij} + \mathbb{A}_{ij}$, where the first term is the isotropic, Heisenberg-like damping [$\alpha = \frac{1}{3}\textrm{Tr}(A) $], the second term is the anisotropic, Ising-like damping [$\textrm{Tr}(\mathbb{I}) =0$] and $\mathbb{A}_{ij}= \varepsilon_{ijk}\bm{D}_k$ is the chiral damping. For sake of brevity we define $\mathcal{A} = \alpha \mathbb{1} + \mathbb{I} $.}
The transformation of the full magnetization dynamics given in Eq.\ (\ref{eq:motion_full_final1}) to the LL form leads to (see Appendix B for details) 
\begin{widetext}
\begin{align}
 \left(\Psi ^2 + \mathbb{F}\right)\cdot  \frac{\partial\bm{M}}{\partial t}
 = & \underbrace{-\gamma_0\Psi\bm{M}\times\bm{H}_{\rm eff} -\gamma_0\bm{M}\times\left[\mathcal{A}\cdot \left(\bm{M}\times\bm{H}_{\rm eff}\right)\right]}_{\rm Landau-Lifshitz\,\, equations} 
 - \underbrace{\gamma\Psi \bm{M}\times \bm{B}_{\rm opt} - \gamma \bm{M}\times\left[\mathcal{A} \cdot \left(\bm{M}\times \bm{B}_{\rm opt}\right)\right]}_{\rm optical\,\, spin-orbit \,\,torques}\nonumber\\
& - \underbrace{\frac{\Psi}{e}\big(\bm{j}_{\rm NR}\cdot\bm{\nabla}\big)\bm{M} - \frac{1}{e}\bm{M}\times \left[\mathcal{A} \cdot \big(\bm{j}_{\rm NR}\cdot\bm{\nabla}\big)\bm{M}\right]}_{\rm current-induced \,\,STT} 
  - \underbrace{\frac{\Psi\xi}{e} \Big[\bm{M}\big(\bm{E}\cdot(\bm{\nabla}\times\bm{M}) \big)+\big((\bm{M}\times\bm{E})\cdot\bm{\nabla}\big)\bm{M}\Big] }_{\rm field-induced\,\, even\,\,SOT}\nonumber\\
 & - \underbrace{\frac{\xi}{e}\bm{M}\times\left\{\mathcal{A} \cdot  \Big[\bm{M}\big(\bm{E}\cdot(\bm{\nabla}\times\bm{M}) \big)+\big((\bm{M}\times\bm{E})\cdot\bm{\nabla}\big)\bm{M}\Big]\right\}}_{\rm field-induced\,\,odd\,\, SOT} \,,
 \label{final_equation}
\end{align}
\end{widetext}
with the general form of the tensor 
$\mathbb{F} = \alpha^2M^2\mathbb{1} - (\alpha\mathbb{1} + \mathbb{I})(\bm{M}\cdot \mathbb{I}\cdot \bm{M}) + (\bm{M}\cdot \mathbb{I})^2 - M^2\mathbb{I}^2 + \bm{M}(\bm{M}\cdot \mathbb{I}^2)$ and $\Psi = 1+\bm{M}\cdot\bm{D}$ which serves as a normalization parameter. 
One can recognize that the first two terms on the right-hand side of Eq.\ (\ref{final_equation}) are exactly the LL equation of spin motion with tensorial damping parameter. The additional terms give the influences of the various derived torques.

  A distinction between the LL and LLG magnetization dynamics (Eqs.\ (\ref{final_equation}) and (\ref{eq:motion_full_final1})) that has not be highlighted in earlier investigations can now be recognized. We find that the STT and SOT terms do not appear in exactly the same form in both the LL and LLG equations of spin dynamics. Although previously often the STT and SOT terms have been added simply in the same way to the LL or LLG equation of motion, our derivation suggests that the torque terms should appear in distinct forms.

\section{Discussion}
{In the past several expressions have been proposed for spin-transfer torques caused by spin-polarized currents in ferromagnetic systems with inhomogeneous magnetization distributions \cite{Slonczewski1996,Berger1996,Bazaliy1998,Stiles2002,Fernandez2004,Li2004,Tatara2004,Ansermet2004,Thiaville2005,Ralph2008,Brataas2012,Bijl2012,Kim2013,Cheng2013,Kim2015}. The derivations have been based on various considerations, such as the ``\textit{s-d}'' model, total electro-magnetic free energy, spin-wave theory, spin conservation, symmetry considerations, and micromagnetic simulations. These investigations have identified the adiabatic Berger STT, $T_{\rm STT} \propto (j_{\rm NR} \cdot \bm{\nabla}) \bm{M}$ \cite{Berger1996}, which can be seen as the continuous limit of the Slonczewski spin-transfer torque \cite{Slonczewski1996} for infinitely thin successive layers. 

The nonadiabatic STT term, 
$\bm{M} \times (\bm{j}_{\rm NR} \cdot \bm{\nabla} ) \bm{M} $ (second term on 2$^{\rm nd}$ line of Eq.\ (\ref{final_equation})), has been obtained previously,  by Zhang and Li \cite{Zhang2004} and others \cite{Thiaville2005,Kim2013,Cheng2013,Kim2015}.
The need for this term has been stressed by micromagnetic simulations \cite{Thiaville2005,Beach2008}. These simulations introduced a so-called nonadiabaticity parameter $\beta$ that defines the relative strengths of the nonadiabatic and adiabatic STTs, whose value was estimated to be comparable to the Gilbert damping $\alpha$ \cite{Thiaville2005}. Our Dirac theory derivation 
gives that $\beta$ is equal to $\mathcal{A}$ (using that $\Psi = 1$ for isotropic damping).
 We note however that these relations are derived here for the \textit{intrinsic} Gilbert damping and nonadiabaticity parameter. Additional damping effects, as e.g., nonlocal damping and spin pumping contributions could change the \textit{effective} Gilbert damping \cite{Tserkovnyak2002,Tserkovnyak2009,Nembach2013,Wang2015}, thus also making it different from $\beta$. In our case,
the nonadiabaticity parameter can be accounted for through the Gilbert damping parameter because the electronic structure expressions for the damping parameter contain the broadening that is the relaxation time (see e.g., Refs.\ \cite{Mondal2016} and \cite{Mondal2017Nutation} for expressions). 

 The optical spin-orbit torques in Eq.\ (\ref{final_equation}) have the same form as those of the precession and damping in the standard LL equation. However, the first optical spin-orbit torque (odd in magnetization) is already of relativistic origin through the relativistic optomagnetic field, and the second optical spin-orbit torque (even in magnetization) is of higher order in the relativistic effects because it contains the optomagnetic field (which is relativistic by our definition) as well as the Gilbert damping parameter. 
 
 The remaining  terms in Eq.\ (\ref{final_equation})  (and Eq.\ (\ref{eq:motion_full_final1})) arise as SOTs due to the spin-orbit coupling. It deserves to be noted that for these SOTs a relativistic derivation on the basis of the Dirac theory is indispensible. While identical forms of the nonrelativistic STTs can be derived within various theories, for the SOTs the precise form of the spin-orbit related terms in the Hamiltonian is essential. One term, the second term in the field-induced even SOT in Eq.\ (\ref{final_equation}), was proposed earlier on the basis of symmetry considerations \cite{Bijl2012}.
 
We observe that there exist two different SOTs in Eq.\ (\ref{final_equation}): (i) the even-in-magnetization - even SOT, and (ii) the odd-in-magnetization - odd SOT. Both of them account for the inhomogeneity of the magnetization.  Previous investigations \cite{Miron2011,Gambardella2011,Garello2013,Haney2013,Kurebayashi2014,Lee2015} have identified two contributions to the SOT, one having fieldlike form,  being odd-in-$M$, 
$T_{\rm SOT}^{\rm odd} = t^{\rm odd} \bm{M} \times (\hat{\bm{z}} \times {\bm{E}})$ and one having dampinglike form, being even-in-$M$,
$T_{\rm SOT}^{\rm even} = t^{\rm even} \bm{M} \times [(\hat{\bm{z}} \times {\bm{E}}) \times {\bm{M}}]$, where $\hat{\bm{z}}$ is the unit vector perpendicular to the interface or surface  and the $E$-field or current is typically in the $xy$-plane. These SOTs are often discussed phenomenologically in the literature in the case of surfaces or interfaces having Rashba-like couplings which are derived from zeroth order in the gradient of magnetization \cite{Rashba1984,Edelstein1990,Bijl2012}
 where the Rashba-like field is expressed as $\hat{\bm{z}}\times\bm{E}$.
However, these expressions do not account for the magnetization inhomogeneity, even though they are field-induced. Our formulation has the advantage that it provides an explicit derivation that leads to general expressions for the SOTs that account for inhomogeneous magnetization.

Several recent studies have investigated SOTs without assuming a special form of spin-orbit interaction. In particular, \textit{ab initio} calculations of SOTs have been performed for several magnetic metal -- nonmagnetic metal bilayers and trilayers
\cite{Freimuth2014PRB,Freimuth2015,Wimmer2016,Mahfouzi2018}.  These investigations provide microscopic information on the size and origin of the SOT at the interfaces, whereas our formulation provides the relativistic functional dependence of the SOTs on the electric field and inhomogeneous magnetization. In addition, our equations are directly suitable for  micromagnetic simulations.

In our derivation, the first and third  and fourth terms of the second line in Eq.\ (\ref{final_equation}) cannot be treated as a simple torque as the torque has to be perpendicular to the magnetization. The third and fourth terms in the second line and the torques in the last line 
of Eq.\ (\ref{final_equation}) provides the SOT in the adiabatic and nonadiabatic limit, respectively. 
As these SOTs derive from the spin-orbit coupling, they explain the spin-torques due to the Hall effect \citep{Bijl2012}. This means that the current and the $E$-field direction have to be perpendicular to each other.
It should also be observed that the even-SOT torques depend on the chiral damping  $\bm{D}$ through the parameter $\Psi$, therefore they could play an important role to manipulate magnetic textures. 
The terms in the last line of Eq.\ (\ref{final_equation}) provide the nonadiabatic SOT as the relaxation times are accounted for through the Gilbert damping tensor, $\mathcal{A}$. At the same time, these torques are odd in the  magnetization. For a scalar Gilbert damping parameter, the first term vanishes, however, for tensorial damping it does not vanish.
The second term, on the other hand, can be seen as a nonadiabatic SOT which is odd in magnetization. 
 The odd SOTs do not depend on the chiral damping $\Psi$.  A fundamental difference between the here-derived even and odd SOTs is that the even SOT is first order relativistic (scales with $1/c^2$), however, the odd SOT is a higher order relativistic torque (scales with $1/c^4$).
 
Finally, we make a comparison between the here-derived spin torques and the possible forms of spin torques predicted on the basis of symmetry considerations in the investigation by van der Bijl and Duine \cite{Bijl2012}. They considered all possible torques that are first-order linear in the gradient of the magnetization and applied field $E$. In this way they identified $(\bm{E}\cdot \bm{\nabla})\bm{M}$ and $\bm{M} \times (\bm{E} \cdot \bm{\nabla})\bm{M}$, the equivalents of the current-induced STT torques. They further obtained $ [(\bm{M} \times \bm{E}) \cdot \bm{\nabla}] \bm{M}$, the field-induced even SOT, which they, consistent with our relativistic derivation, ascribe to the Hall current. They also proposed several other possible torque forms, which however are not given by our derivation. Conversely, our first term of the field-induced even SOT in  Eq.\ (\ref{eq:motion_full_final1}) could not be obtained by them because it does not have pure torque form. Apart from these spin torques, we have derived the optical spin-orbit torques which are higher-order than linear in the applied $E$-field and have thus not been considered by van der Bijl and Duine  \cite{Bijl2012}.

\section{Conclusions}
Starting from the Dirac-Kohn-Sham theory we have derived the full magnetization dynamics with the effects of spin-polarized currents included. We have shown that the  nonrelativistic and relativistic contributions to the dynamics can be separated and we have derived their distinct magnetization dynamics contributions. 
Our results show that the nonrelativistic magnetization dynamics consist of Larmor precession and the adiabatic STT. On the other hand, the relativistic magnetization dynamics is more complicated. The latter consists of relativistic counterparts of precession, Gilbert damping with a tensorial dissipation parameter, optical spin-orbit torque, the nonadiabatic STT, and adiabatic and nonadiabatic SOTs that derive from current density due to spin-orbit coupling. These SOTs take into account the inhomogeneity in the magnetization and are $E$-field induced.

The current-induced nonrelativistic torque explains the 
Berger STT within the adiabatic limit. The corresponding current is unidirectional and perpendicular to the surface or interfaces in a multilayered system, thus it can explain several STT-induced phenomena. On the other hand the spin-orbit current depends on the electric field and the magnetization, limiting the current to lie within a plane that is perpendicular to the nonrelativistic current. This relativistic current gives a spin-orbit torque on the magnetization. 
 We have shown that the here-derived SOTs are field-induced and,  in contrast to previous expressions, take into account the magnetization inhomogeneity in a preferred plane (not necessarily the surface or interfaces).  The derived expressions therefore describe a more general form of the SOTs.

In the later part, once we have derived the full magnetization dynamics including the nonrelativistic and relativistic contributions, we  have transformed the LLG form of the magnetization dynamics to the LL form. By doing so, we re-derive the nonadiabatic torques as well.  We obtain the Zhang-Li nonadiabatic STT,
 and show that the nonadiabaticity parameter $\beta$ strongly depends on the  intrinsic Gilbert damping parameter and scales 
linearly. 
 We have further obtained an optical spin torque, which, 
if high enough, can lead to a spin switching.
Consequently, we have derived the full magnetization dynamics with the effects of spin-polarized currents included within a fundamental framework, and within this unified framework we  have derived  all the dynamical processes i.e., magnetization precession, damping, optical spin-orbit torque, adiabatic and nonadiabatic spin-transfer torque, adiabatic and nonadiabatic spin-orbit torques. 

\begin{acknowledgments}

 We thank U. Nowak, U.\ Ritzmann and A. Aperis for valuable discussions.
 This work has been supported by the Swedish Research Council (VR), the Knut and Alice Wallenberg Foundation (Contract No.\ 2015.0060), the European Union's Horizon2020 Research and Innovation Programme under grant agreement No.\ 737709 (FEMTO-TERABYTE, {\blue http://www.physics.gu.se/femtoterabyte}).

\end{acknowledgments}

\appendix

\begin{widetext}

\section{Landau-Lifshitz spin dynamics for a general time-dependent field}
For a general time-dependent magnetic field, we have recently shown that the traditional Landau-Lifshitz-Gilbert equation of motion is not valid anymore, and we derived an improved LLG equation that includes the {\it field-derivative torque}, as given by the expression \cite{Mondal2016}
\begin{align}
\label{newLLG}
	\frac{\partial \bm{M}}{\partial t} & = -\gamma_0\bm{M}\times \bm{H}_{\rm eff} + \bm{M} \times \left[\bar{A}\cdot \left(\frac{\partial\bm{M}}{\partial t}+\frac{\partial\bm{H}}{\partial t}\right)\right]\,,
\end{align} 
where {$\bm{M} \times \frac{\partial\bm{H}}{\partial t}$ is the field-derivative torque,} and the following equation for the damping tensor has been obtained
\begin{align}
	\bar{A}_{ij} & = -\frac{e\mu_0}{8m^2c^2}\sum_{n}\langle r_ip_j + p_jr_i\rangle - \langle r_np_n +p_nr_n \rangle \delta_{ij}\,.
\end{align} 
{Note that, due to the additional term $\partial \bm{H} / \partial t$ in the damping term, $ \bar{A} \neq A$.}
For $i\neq j$, the different components of $r_i$ and $p_j$ commute with each other and thus the matrix $\bar{A}_{ij}$ is symmetric. Now we split the damping tensor in two parts: $\bar{A}_{ij} = \alpha \delta _{ij} + \Lambda_{ij}$, where $\alpha = \frac{1}{3}{\rm Tr} (\bar{A}_{_{ij}})$. 
Therefore the trace of the symmetric matrix {$\Lambda_{ij}$} 
will be zero. Equation (\ref{newLLG}) is then rewritten as
	\begin{align}
	\label{newLLG1}
	\frac{\partial \bm{M}}{\partial t}  = & -\gamma_0\bm{M}\times \bm{H}_{\rm eff} + \alpha \bm{M} \times  \left(\frac{\partial\bm{M}}{\partial t}+\frac{\partial\bm{H}}{\partial t}\right) + \bm{M} \times \left[\Lambda\cdot \left(\frac{\partial\bm{M}}{\partial t}+\frac{\partial\bm{H}}{\partial t}\right)\right]\,.
\end{align} 
To evaluate the last two terms in Eq.\ (\ref{newLLG1}), we perform a suitable vector multiplication,
\begin{align}
	\bm{M}\times \left(\frac{\partial \bm{M}}{\partial t} + \frac{\partial \bm{H}}{\partial t} \right) 
	 =&  -\gamma_0 \bm{M}\times \left(\bm{M}\times \bm{H}_{\rm eff}\right) +\alpha \bm{M} \left(\bm{M}\cdot \frac{\partial\bm{H}}{\partial t}\right) -\alpha M^2 \left(\frac{\partial \bm{M}}{\partial t} + \frac{\partial \bm{H}}{\partial t} \right)\nonumber\\
	& + \bm{M} \left[\bm{M} \cdot \left\{\Lambda\cdot \left(\frac{\partial\bm{M}}{\partial t}+\frac{\partial\bm{H}}{\partial t}\right)\right\}\right] - M^2 \left[\Lambda\cdot\left(\frac{\partial\bm{M}}{\partial t}+\frac{\partial\bm{H}}{\partial t}\right) \right] + \bm{M}\times \frac{\partial \bm{H}}{\partial t}\,,
\end{align} 
and similarly,
\begin{align}
	\bm{M} \times \left[\Lambda\cdot \left(\frac{\partial\bm{M}}{\partial t}+\frac{\partial\bm{H}}{\partial t}\right)\right] & = -\gamma_0\bm{M}\times\left[ \Lambda\cdot\left(\bm{M}\times \bm{H}_{\rm eff}\right)\right]+ \alpha\bm{M}\times \left[\Lambda\cdot \left\{\bm{M} \times  \left(\frac{\partial\bm{M}}{\partial t}+\frac{\partial\bm{H}}{\partial t}\right)\right\}\right] \nonumber\\
	& +\bm{M}\times \Lambda\cdot\left\{ \bm{M} \times \left[\Lambda\cdot \left(\frac{\partial\bm{M}}{\partial t}+\frac{\partial\bm{H}}{\partial t}\right)\right]\right\}\,.
\end{align}
Now, following the same procedure as described in Ref.\ \cite{Mondal2016} we arrive at 
\begin{align}
	\left(\mathbbm{1} + \mathbbm{G}\right) \cdot \left(\frac{\partial \bm{M}}{\partial t} + \frac{\partial\bm{H}}{\partial t}\right)  = & -\gamma_0 (\bm{M}\times \bm{H}_{\rm eff})  - \gamma_0\,   \bm{M}\times \left[(\alpha \mathbbm{1}+\Lambda) \cdot (\bm{M}\times \bm{H}_{\rm eff})\right] \nonumber\\ 
	& + \left[\alpha^2\bm{M} \,\mathbbm{1} -\alpha (\bm{M}\cdot \Lambda)\right]  \left(\bm{M}\cdot \frac{\partial\bm{H}}{\partial t}\right) + \alpha \bm{M}\times \frac{\partial \bm{H}}{\partial t} + \frac{\partial\bm{H}}{\partial t}\,,
	\label{fullLL}
\end{align}
where the newly derived tensor $\mathbbm{G}$ depends on the magnetization and is represented as (see e.g., Eq.\ (B11) of Ref.\ \cite{Mondal2016})
\begin{align}
	\mathbbm{G} & = \alpha^2M^2\mathbbm{1}  - (\alpha\mathbbm{1}+\Lambda) [\bm{M} \cdot \Lambda \cdot \bm{M}] - M^2\Lambda^2 + (\bm{M}\cdot \Lambda)^2 + \bm{M} (\bm{M}\cdot \Lambda^2)\,.
\end{align}
Equation (\ref{fullLL}) is the Landau-Lifshitz equation of spin motion for a general time-dependent field. Note that if $\bm{H} = 0$, the dynamics collapses to the original LL equation. Although in general the damping is tensorial, it is of interest (e.g., for numerical simulations) to derive a compact form of the LL equation of motion for a general field yet for isotropic, scalar damping. Thus, taking $\Lambda = 0$, the concrete form of the LL equation of spin motion with applied general field and scalar damping will be 
\begin{align}
	\left(1 + \alpha^2M^2 \right) \frac{\partial \bm{M}}{\partial t}    = -\gamma_0 \bm{M}\times \left(\bm{H}_{\rm eff} - \frac{\alpha}{\gamma_0}\frac{\partial \bm{H}}{\partial t} \right)  - \gamma_0\,\alpha   \bm{M}\times  \left[\bm{M}\times \left(\bm{H}_{\rm eff} - \frac{\alpha}{\gamma_0}\frac{\partial \bm{H}}{\partial t} \right) \right]\,,
\end{align} 
where the field-derivative torque appears in the spin precession and in the spin dissipation terms.

\section{Landau-Lifshitz equation with relativistic and nonrelativistic spin torques}

In Sec.\ \ref{sec:fulldyn} we derived the full LLG magnetization dynamics including the nonrelativistic and relativistic spin-current contributions; the equation of motion  was  found to be given by 
\begin{align}
\frac{\partial\bm{M}}{\partial t} & = - \gamma_0\bm{M}\times\bm{H}_{\rm eff} +\bm{M}\times\Big( A \cdot\frac{\partial\bm{M}}{\partial t}\Big) - \gamma \bm{M}\times \bm{B}_{\rm opt} - \frac{1}{e}\big(\bm{j}_{\rm NR}\cdot\bm{\nabla}\big)\bm{M}+\frac{\xi}{e} \Big[\bm{M}\big(\bm{E}\cdot(\bm{\nabla}\times\bm{M}) \big)+\big((\bm{M}\times\bm{E})\cdot\bm{\nabla}\big)\bm{M}\Big]\,.
\label{full-dynamics}
\end{align}
In its most general form the Gilbert damping $A_{ij}$ is a tensor. This tensor can be decomposed as {$ A_{ij}= \alpha \delta_{ij} + \mathbb{I}_{ij} + \mathbb{A}_{ij}$, where the first two terms are its symmetric part and the third term is its antisymmetric part.} {The symmetric part can be decomposed in the isotropic, Heisenberg-like damping [$\alpha = \frac{1}{3}\textrm{Tr}(A^{\rm sym}) $], and an anisotropic, Ising-like damping [$\textrm{Tr}(\mathbb{I}) =0$]. However, the antisymmetric part leads to a chiral, Dzyaloshinskii-Moriya-like damping represented by a corresponding vector $\bm{D}$, as  $\mathbb{A}_{ij}= \varepsilon_{ijk}\bm{D}_k$, with $\varepsilon$ the Levi-Civita tensor.}
The spin dissipation dynamics can {then} be expressed as 
\begin{align}
\bm{M}\times\Big( A \cdot\frac{\partial\bm{M}}{\partial t}\Big)  =  \alpha\bm{M}\times\frac{\partial\bm{M}}{\partial t} + \bm{M}\times\left[ \mathbb{I} \cdot\frac{\partial\bm{M}}{\partial t}\right] + \bm{M}\times\left[\bm{D}\times \frac{\partial\bm{M}}{\partial t}\right]\,.
\end{align}
As shown previously \cite{Mondal2016}, the Dzyaloshinskii-Moriya-like damping contributes to the renormalization of the dynamics and therefore, the spin dynamics in Eq.\ (\ref{full-dynamics}) can be written as
\begin{align}
\Psi \frac{\partial\bm{M}}{\partial t}  = & - \gamma_0\bm{M}\times\bm{H}_{\rm eff} +\alpha\bm{M}\times\frac{\partial\bm{M}}{\partial t} + \bm{M}\times\left[ \mathbb{I} \cdot\frac{\partial\bm{M}}{\partial t}\right] -\gamma \bm{M}\times \bm{B}_{\rm opt} \nonumber\\
& - \frac{1}{e}\big(\bm{j}_{\rm NR}\cdot\bm{\nabla}\big)\bm{M} - \frac{\xi}{e} \Big[\bm{M}\big(\bm{E}\cdot(\bm{\nabla}\times\bm{M}) \big)+ \big((\bm{M}\times\bm{E})\cdot\bm{\nabla}\big)\bm{M}\Big]\,,
\label{full-dyna1}
\end{align}
where we define $\Psi = 1 + \bm{M}\cdot \bm{D}$.

Next, our objective is to obtain the Landau-Lifshitz form of spin dynamics with spin-current torques.
To this end we follow the usual procedure to derive the LL spin dynamics, i.e., we have to calculate the second term of Eq.\ (\ref{full-dyna1}). To do so, we take a cross product with the magnetization on the both sides of the last equation,
\begin{align}
\Psi \bm{M}\times \frac{\partial\bm{M}}{\partial t}  = & - \gamma_0\bm{M}\times \left(\bm{M}\times\bm{H}_{\rm eff}\right) -\alpha M^2\,\frac{\partial\bm{M}}{\partial t} - M^2 \left[ \mathbb{I} \cdot\frac{\partial\bm{M}}{\partial t}\right] + \bm{M}\left(\bm{M}\cdot\left[ \mathbb{I} \cdot\frac{\partial\bm{M}}{\partial t}\right] \right) \nonumber\\
& - \gamma \bm{M}\times \left(\bm{M}\times \bm{B}_{\rm opt}\right) 
 - \frac{1}{e}\bm{M}\times \left[\big(\bm{j}_{\rm NR}\cdot\bm{\nabla}\big)\bm{M}\right] - \frac{\xi}{e}\bm{M}\times \Big[\bm{M}\big(\bm{E}\cdot(\bm{\nabla}\times\bm{M}) \big)+ \big((\bm{M}\times\bm{E})\cdot\bm{\nabla}\big)\bm{M}\Big]\,.
\end{align}
However, the calculation of the third term on the right-hand side of Eq.\ (\ref{full-dyna1}) involves much more. 
We calculate this term as, 
\begin{align}
\Psi \bm{M}\times \left[\mathbb{I}\cdot\frac{\partial\bm{M}}{\partial t} \right]  = &- \gamma_0\bm{M}\times\left[\mathbb{I}\cdot \left(\bm{M}\times\bm{H}_{\rm eff}\right)\right] + \alpha M^2\left[ \mathbb{I} \cdot\frac{\partial\bm{M}}{\partial t}\right] + \alpha (\bm{M}\cdot \mathbb{I}\cdot \bm{M})\frac{\partial\bm{M}}{\partial t} - \alpha \bm{M}\left[ \bm{M}\cdot \left(\mathbb{I}\cdot \frac{\partial\bm{M}}{\partial t}\right)\right]\nonumber\\
& + (\bm{M}\cdot \mathbb{I}\cdot \bm{M}) \left(\mathbb{I}\cdot \frac{\partial\bm{M}}{\partial t}\right) - (\bm{M}\cdot \mathbb{I})\left[ \bm{M}\cdot \left(\mathbb{I}\cdot \frac{\partial\bm{M}}{\partial t}\right)\right] + M^2 \left(\mathbb{I}^2\cdot \frac{\partial\bm{M}}{\partial t}\right) - \bm{M}\left[\bm{M}\cdot \left(\mathbb{I}^2\cdot \frac{\partial\bm{M}}{\partial t}\right)\right]\nonumber\\
& -\gamma \bm{M}\times\left[\mathbb{I}\cdot \left(\bm{M}\times \bm{B}_{\rm opt}\right)\right] - \frac{1}{e}\bm{M}\times \left[\mathbb{I} \cdot \big(\bm{j}_{\rm NR}\cdot\bm{\nabla}\big)\bm{M}\right]\nonumber\\
&  - \frac{\xi}{e}\bm{M}\times\left\{\mathbb{I} \cdot  \Big[\bm{M}\big(\bm{E}\cdot(\bm{\nabla}\times\bm{M}) \big)+ \big((\bm{M}\times\bm{E})\cdot\bm{\nabla}\big)\bm{M}\Big]\right\}\,.
\end{align}
Next, we substitute these last two equations back into Eq.\ (\ref{full-dyna1}) and obtain the full magnetization dynamics with spin-current torques in LL form that can be written as
\begin{align}
\left(\Psi ^2 + \mathbb{F}\right)\cdot  \frac{\partial\bm{M}}{\partial t}  = & -\gamma_0\Psi\bm{M}\times\bm{H}_{\rm eff} -\gamma_0\bm{M}\times\left[(\alpha\mathbb{1} + \mathbb{I})\cdot \left(\bm{M}\times\bm{H}_{\rm eff}\right)\right] 
 -\gamma \Psi \bm{M}\times \bm{B}_{\rm opt} - \gamma \bm{M}\times\left[(\alpha\mathbb{1} + \mathbb{I})\cdot \left(\bm{M}\times \bm{B}_{\rm opt}\right)\right]\nonumber\\
& - \frac{\Psi}{e}\big(\bm{j}_{\rm NR}\cdot\bm{\nabla}\big)\bm{M} - \frac{1}{e}\bm{M}\times \left[(\alpha\mathbb{1} + \mathbb{I}) \cdot \big(\bm{j}_{\rm NR}\cdot\bm{\nabla}\big)\bm{M}\right]
 -\frac{\Psi\xi}{e} \Big[\bm{M}\big(\bm{E}\cdot(\bm{\nabla}\times\bm{M}) \big)+ \big((\bm{M}\times\bm{E})\cdot\bm{\nabla}\big)\bm{M}\Big] \nonumber\\
 & -\frac{\xi}{e}\bm{M}\times\left\{(\alpha\mathbb{1} + \mathbb{I}) \cdot  \Big[\bm{M}\big(\bm{E}\cdot(\bm{\nabla}\times\bm{M}) \big)+ \big((\bm{M}\times\bm{E})\cdot\bm{\nabla}\big)\bm{M}\Big]\right\} \,,
\end{align}
where the  tensor $\mathbb{F}$ is given by
$\mathbb{F} = \alpha^2M^2\mathbb{1} - (\alpha\mathbb{1} + \mathbb{I})(\bm{M}\cdot \mathbb{I}\cdot \bm{M}) + (\bm{M}\cdot \mathbb{I})^2 - M^2\mathbb{I}^2 + \bm{M}(\bm{M}\cdot \mathbb{I}^2)$, while considering the fact that ${\rm Tr}(\mathbb{I}_{ij}) = 0$ here. 
Lastly, we note that if one for sake of simplicity considers only the scalar Gilbert damping parameter (i.e., $ \mathbb{I} =0$, $\bm{D}=0$), the resulting spin dynamics resembles the LL equation of motion but including the additional current and field-induced torque terms,
\begin{align}
(1+\alpha^2M^2)\frac{\partial\bm{M}}{\partial t}  = & - \gamma_0\bm{M}\times\bm{H}_{\rm eff} - \alpha\gamma_0\bm{M}\times (\bm{M}\times\bm{H}_{\rm eff}) -\gamma \bm{M}\times \bm{B}_{\rm opt} - \gamma \alpha \bm{M}\times\left(\bm{M}\times \bm{B}_{\rm opt}\right) \nonumber\\
& - \frac{1}{e}\big(\bm{j}_{\rm NR}\cdot\bm{\nabla}\big)\bm{M} - \frac{\alpha}{e}\bm{M}\times\left[\big(\bm{j}_{\rm NR}\cdot\bm{\nabla}\big)\bm{M}\right]\nonumber\\
& -\frac{\xi}{e} \Big[\bm{M}\big(\bm{E}\cdot(\bm{\nabla}\times\bm{M}) \big)+ \big((\bm{M}\times\bm{E})\cdot\bm{\nabla}\big)\bm{M}\Big] - \frac{\alpha\xi}{e}\bm{M}\times \left[ \big((\bm{M}\times\bm{E})\cdot\bm{\nabla}\big)\bm{M}\right]. 
\end{align}
\end{widetext}

\bibliographystyle{apsrev4-1}

\begin{thebibliography}{116}
\bibitem [{not()}]{note1}%
  \BibitemOpen
  \href@noop {} {}\bibinfo {note} {Note that when the momentum operator and a
  (vector) function are enclosed in round brackets the momentum operator acts
  only on this function.}\BibitemShut {Stop}%
\end{thebibliography}

%
\end{document}